\documentclass{jfm}

\usepackage{graphicx}
\usepackage{newtxtext}
\usepackage{newtxmath}
\usepackage{natbib}
\usepackage{hyperref}
\hypersetup{
    colorlinks = true,
    urlcolor   = blue,
    citecolor  = black,
}

\graphicspath{{figures/}} 

\shorttitle{Bubble collapse near porous plates}
\shortauthor{E. D. Andrews, D. Fern\'andez Rivas and I. R. Peters}

\title{Bubble collapse near porous plates}

\author{Elijah D. Andrews\corresp{\email{e.d.andrews@soton.ac.uk}}\aff{1},
 David Fern\'andez Rivas\aff{2}
 \and Ivo R. Peters\aff{1}}

\affiliation{\aff{1}Faculty of Engineering and Physical Sciences, University of Southampton, Southampton SO17 1BJ, UK
\aff{2}Mesoscale Chemical Systems Group, MESA+ Institute, TechMed Centre and Faculty of Science and Technology, University of Twente,
P.O. Box 217, 7500 AE Enschede, The Netherlands}
\begin{document}

\maketitle

\begin{abstract}
The collapse of a gas or vapour bubble near a non-porous boundary is directed at the boundary due to the asymmetry induced by the nearby boundary. High surface pressure and shear stress from this collapse can damage, or clean, the surface. A porous boundary, such as a filter, would act similarly to a non-porous boundary but with reduced asymmetry and thus reduced effect. Prior research has measured the cleaning effect of bubbles on filters using ultrasonic cleaning, but it is not known how the bubble dynamics are fundamentally affected by the porosity of the surface. We address this question experimentally by investigating how the standoff distance, porosity, pore size, and pore shape affect two collapse properties: bubble displacement and bubble rebound size. We show that these properties depend primarily on the standoff distance and porosity of the boundary and extend a previously developed numerical model that approximates this behaviour. Using the numerical model in combination with experimental data, we show that bubble displacement and bubble rebound size each collapse onto respective single curves.
\end{abstract}

\section{Introduction}
\label{sec:introduction}
Collapsing bubbles can be found in numerous physical systems. Typically this involves bubbles collapsing in proximity to various boundaries. The high surface pressures and shear stresses generated by collapsing bubbles can damage, or clean, the boundary. This cleaning effect can be harnessed through processes such as ultrasonic cleaning \citep{ Reuter2016} which can involve complex geometries \citep{Verhaagen2016a}.

Single bubbles collapsing near simple geometries, such as flat rigid boundaries, have been widely investigated \citep{Benjamin1966}. Much of this research has focused on understanding bubble morphology and jetting \citep{Kroninger2010, Zhang2019} and the effect of bubble collapse on the boundary. Of particular importance for consideration of cleaning and damage are measurements of nearby surface shear stress and pressure, both experimentally \citep{Dijkink2008a, Luo2018, Occhicone2019} and numerically \citep{Li2016, Koukouvinis2018, Zeng2018}. The resulting cleaning and surface damage have also been widely investigated \citep{Ohl2006, Sagar2020, Reuter2022}.

There has been a recent effort to characterise single bubble collapse in a range of complex geometries. For example, characterising jet direction in a selection of geometries including in concave corners \citep{PhysRevFluids.3.081601}, inside rectangular and triangular prisms \citep{Molefe2019}, above slots \citep{Andrews2020}, and in the corner of a wall and a free surface \citep{Kiyama2021}. Bubble morphology, flow properties, and jetting behaviours have also been investigated in combinations of concave corners and free surfaces \citep{Zhang2017, Brujan2018}, between two parallel rigid boundaries \citep{Brujan2019, Rodriguez2022}, inside a slot \citep{Brujan2022}, on a convex corner \citep{Zhang2020}, on a crevice \citep{Trummler2020}, and on ridge-patterned structures \citep{Kim2020, Kadivar2022}. Nevertheless, there remain fundamental complex geometries that have yet to be explored.


Porous materials are a large family of complex geometries with a wide range of applications. Broadly, they could be categorised as connected or unconnected. One example of an unconnected porous material is a bed of sand, such as those investigated by \citet{Sieber2022}. The sand creates pores which are permeated by water, but the grains are able to separate which allows for significant deformation of the bed. The influence of the bubble on the boundary is shown to depend on the granular size of the sand, including behaviors such as granular jets and displacement of the boundary material. However, the boundary is modelled as a liquid-liquid interface and it is shown that the bubble rebound and displacement are principally driven by the density difference between the pliable granular suspension and the water.

Connected porous materials have a wide range of applications, of which one very practical application is filters. Filters are typically porous materials through which a fluid is passed in order to remove contamination. This contamination builds up on the filter, reducing performance. Ultrasonic cleaning can be applied to filters in order to remove the built-up contamination from the filter \citep{Reuter2017}.

The simplest example of a porous material is a flat plate with a pattern of through-holes. Some research has investigated the problem of a plate with a through-hole bounding a free surface in relation to breaches of maritime hulls \citep{He2021, Cui2021, Cui2022}. Similarly, \citet{Sun2022} investigated bubble collapse near a rigid surface with a gas-entrapping hole. This concept was also investigated by \citet{Gonzalez-Avila2020a} and \citet{Sun2022a}, who considered surfaces with a pattern of gas-entrapping holes. In all these cases, bubbles were found to translate away from the boundary, acting analogously to a typical free surface.

\cite{Liu2017a} investigated a bubble collapsing between a free surface and a submerged rigid boundary with a through-hole. Bubbles collapsing close to a fully submerged rigid boundary with a single through-hole have also been investigated, both experimentally \citep{Lew2007, Karri2011, Abboud2013} and numerically \citep{Khoo2005}. Similarly, \citet{Moloudi2019} numerically investigated bubble collapse dynamics close to a convex boundary with a through-hole. These investigations revealed similar tendencies of the bubbles during collapse, such as the bubble translating towards and through the holes, the bubble surface expanding into the holes, and stronger counter-jetting than plates without holes.

In this research, we investigate a less-studied phenomenon: how a series of porous plates affect bubble collapse dynamics. We include some comments on bubble morphology and measurements of bubble displacement and bubble rebound radius.

\section{Problem definition}
\label{sec:problem definition}

\begin{figure}
    \centering
    \includegraphics{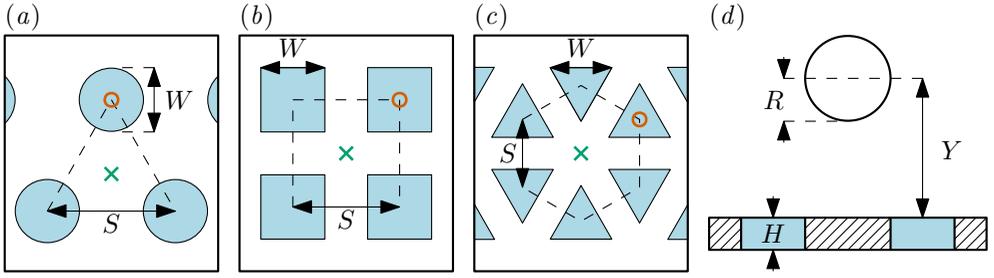}
    \caption{Top-down view schematic diagrams of porous plates with patterns of circular ($a$), square ($b$), or triangular ($c$) holes. The scales in each diagram are arbitrary. Orange circles mark horizontal positions above a hole and green crosses mark horizontal positions between-holes. ($d$) Side-on cross-section view of a porous plate with a bubble positioned vertically above it.}
    \label{fig:schematic}
\end{figure}

We define a porous plate as a thin, rigid plate with a pattern of through-holes. The plate thickness is defined as $H$, with through-holes of characteristic length $W$ and spacing $S$. Three shapes of hole are investigated: circles, squares, and triangles. Circles are the smoothest possible hole shape and triangles are the least-smooth regular polygon, with the tightest corner angle. Squares are somewhere between the two and are used to achieve high void fractions due to their efficient tessellation packing. Schematics of these arrangements are shown in figure \ref{fig:schematic}(\textit{a}-\textit{c}). The void fraction $\phi$ is defined as the fraction of the total plate area that is occupied by holes. In this work, we describe a void fraction $\phi = 0$ as `non-porous' and a void fraction $\phi > 0$ as `porous', both of which are assumed to be rigid. For circular holes the void fraction is
\begin{equation}
    \phi = \frac{\pi \sqrt{3}}{6} \left(\frac{W}{S}\right)^2,
\end{equation}
for square holes the void fraction is
\begin{equation}
    \phi = \left(\frac{W}{S}\right)^2,
\end{equation}
and for triangular holes the void fraction is
\begin{equation}
    \phi = \frac{1}{3} \left(\frac{W}{S}\right)^2.
\end{equation}

Bubbles are positioned vertically at a distance $Y$ from the boundary with time-variant radius $R$ and maximum radius $R_0$. These parameters are shown in figure \ref{fig:schematic}(\textit{d}). Horizontally, the bubble can be positioned in two dimensions with varying proximity to the holes. We identify the two extreme cases as a bubble directly over a hole and a bubble directly over the plate between holes. These two cases represent the minimum and maximum area of solid boundary near the bubble. The horizontal positions of bubbles above holes are shown by the orange circles in figure \ref{fig:schematic}. The horizontal positions of bubbles above the plate between holes are shown by the green crosses in figure \ref{fig:schematic} and are usually referred to hereafter as `between-holes'.

The vertical distance of the bubble is normalised by the maximum bubble radius to produce the dimensionless standoff distance $\gamma = Y / R_0$. The hole size is also normalised with the maximum bubble radius, $W / R_0$.

In this research we investigate fixed, rigid plates with a constant thickness $H = 1~\mathrm{mm}$ and compare only the effects of bubble position and hole geometry.

\section{Methods}
\subsection{Experimental method}
\begin{figure}
    \centering
    \includegraphics{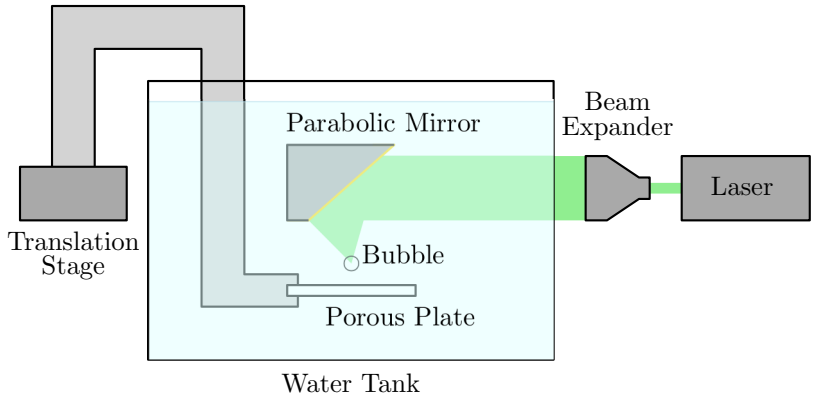}
    \caption{Schematic showing the experimental setup for laser-induced cavitation using an off-axis parabolic mirror to generate bubbles near a porous plate.}
    \label{fig:experiment_diagram}
\end{figure}

Experiments were performed using laser-induced cavitation. Laser-induced cavitation operates by focusing a laser pulse at a point in a volume of water \citep{Lauterborn1972}. The high energy at the focal point forms a plasma which rapidly heats and vaporises the surrounding water, creating a bubble. The focusing optics are typically either a microscope objective or a parabolic mirror. In our previous research, using a microscope objective, we showed that bubbles with greater asymmetry lead to greater spread of experimental measurements of bubble displacement \citep{Andrews2022}. Thus, following the example of \citet{Obreschkow2013}, we have implemented an off-axis parabolic mirror setup, as shown in figure \ref{fig:experiment_diagram}. This method produces more consistent bubbles with improved symmetry, reducing the spread of data. In addition, the larger focusing angle (with a higher equivalent numerical aperture), allows for larger bubbles to be generated without nucleating additional bubbles nearby.

In this research, a Q-switched Nd:YAG laser (`Nano PIV' from Litron Lasers) was used to generate an 8 ns pulse at a wavelength of 532 nm. The laser output energy was approximately 97 mJ and was subsequently attenuated with the attenuator set to between 70 \% and 75 \%. The beam was restricted by an iris to be approximately 3 mm in diameter, in order to remain well-collimated after expansion, and then expanded with a 10x beam expander to approximately 30 mm in diameter. The expanded beam was focused by a gold off-axis parabolic mirror with a focusing angle of approximately 41 degrees (equivalent numerical aperture = 0.35). This methodology produced bubbles with radii ranging from 1.07 mm to 1.65 mm, depending primarily on laser input power and attenuation. Each combination of geometry and position was repeated at least three times. The radius varied with a standard deviation of 0.030 mm and the displacement varied with a standard deviation of 0.033 mm. Both of these values are on a comparable scale to the average measurement resolution of 0.025 mm. Notably, the gold surface absorbs most of the laser energy, but is less likely to degrade when immersed in water when compared to mirror materials designed to operate with a 532 nm laser \citep{Obreschkow2013}. However, a recent study has reported use of an aluminium off-axis parabolic mirror \citep{Sieber2022} so the degradation concern may not be as significant as originally assumed.

\begin{figure}
    \centering
    \includegraphics{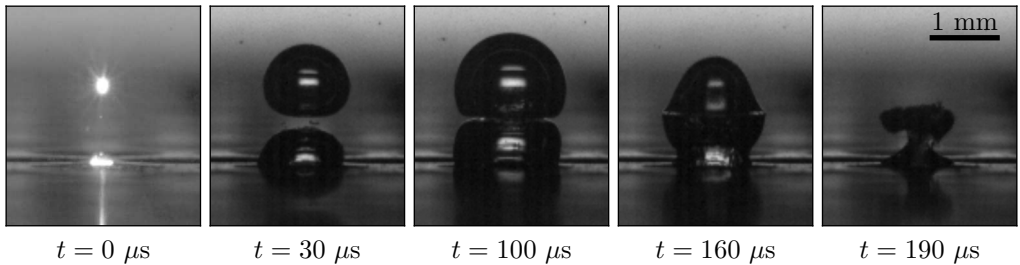}
    \caption{Frames showing the expansion and collapse of two bubbles, one nucleated at the laser focal point and one simultaneously nucleated where the laser impinges on a nearby steel plate. An approximate scale bar is given.}
    \label{fig:surface_nucleation}
\end{figure}

Porous plates were created by laser cutting 50 mm $\times$ 50 mm plates out of stainless steel with a thickness of 1 mm. Stainless steel was selected to ensure rigidity in the plates because other materials, such as 1 mm acrylic, flex significantly under the load of a bubble collapsing, which is known to reduce or reverse displacement \citep{Gibson1982, BRUJAN2001}. Holes had characteristic lengths between 0.40 mm and 4.79 mm, primarily limited at the low end by laser cut quality. Due to the position of the bubble in close proximity to the plate, the diverging laser impinges upon the plate. If the plate is sufficiently close to the focal point, or the focal angle is narrow, the plate can absorb a significant amount of laser energy which can then nucleate a second bubble at the surface of the material. This nucleation is shown in figure \ref{fig:surface_nucleation}. In order to stop significant surface nucleation, the plates were polished to reduce laser energy absorption by increasing the reflectivity. All geometries used are listed in table \ref{tab:all geometries} with measurements of the geometric parameters and range of bubble sizes.

The porous plates were attached to an arm which was moved by a translation stage. The porous plates were submerged in a 174 mm $\times$ 180 mm $\times$ 177 mm glass tank of purified and partially degassed water as shown in figure \ref{fig:experiment_diagram}. The bubbles were back-lit by a 100 W LED panel and recorded at 100 000 frames per second with a high-speed camera (Photron FASTCAM SA-X2 with a 105 mm Nikon Micro-Nikkor lens). A 550 nm longpass filter was used to protect the camera from the laser.

\begin{figure}
    \centering
    \includegraphics{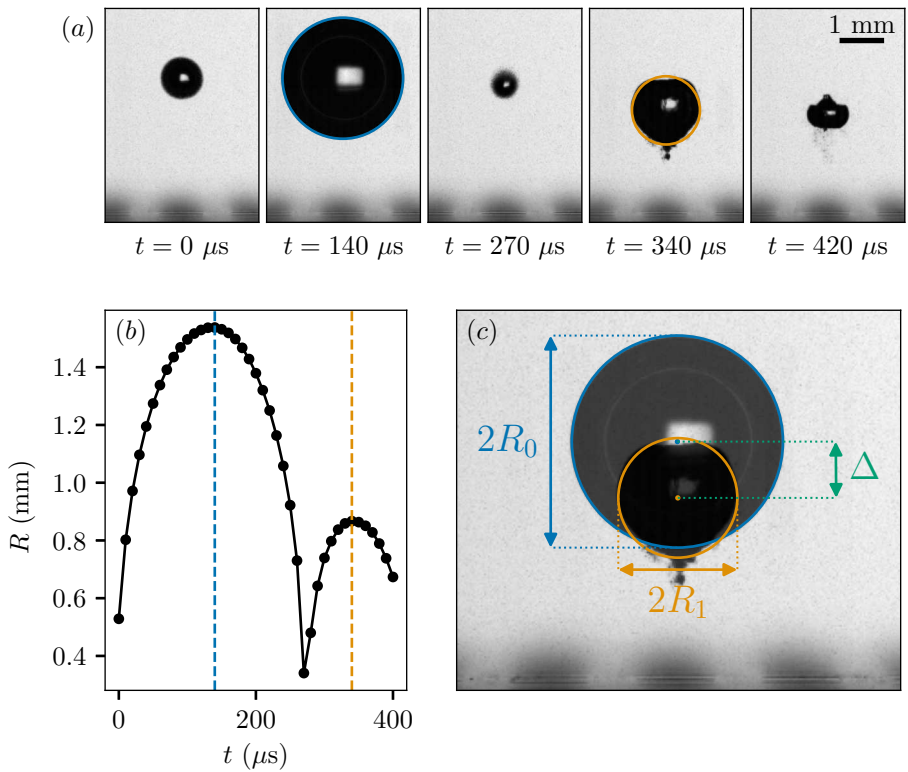}
    \caption{($a$) Frames from a high-speed recording of a bubble expanding and collapsing near a porous boundary. Frames at radius maxima ($t=140~\mathrm{\mu s}$ and $t=340~\mathrm{\mu s}$) feature a circle overlay with area equivalent to the bubble area. ($b$) Bubble radius variation with time. ($c$) Composite image of the two frames recorded at the radius maxima indicated in ($b$) showing measurements of the radius maxima $R_0$ and $R_1$ and bubble displacement $\Delta$.}
    \label{fig:experiment_displacement}
\end{figure}

The experiment produced shadowgraph movies of the bubbles which were then analysed with automatic image analysis software written in Python. For each frame, the background was subtracted, and a binary filter was applied to isolate the bubble. The bubble position was measured as the centroid of the isolated pixels and the bubble radius was calculated as the radius of a circle with equal area to the bubble in the image. For deformed bubbles, where no clear single definition of a radius exists, this is the selected equivalent radius. The three parameters that were extracted from these measurements were the maximum bubble radius $R_0$; the bubble maximum radius after the first rebound $R_1$; and the displacement of the bubble between the two radius maxima $\Delta$. These measurements are shown in figure \ref{fig:experiment_displacement}.

\begin{table}
  \begin{center}
\def~{\hphantom{0}}
  \begin{tabular}{lcccccc}
      Hole shape & $\phi$ (\%)  & $W$ (mm)   &   $A$ (mm$^2$) & $\bar{R_0}$ (mm) \\[3pt]
        non-porous &  0 & - & - & 1.50 \\
        circles &  7.3 & 1.15 & 1.04 & 1.48 \\
        circles & 11.5 & 0.40 & 0.12 & 1.49 \\
        triangles & 13.4 & 1.10 & 0.52 & 1.33 \\
        circles & 14.6 & 1.15 & 1.04 & 1.45 \\
        circles & 21.6 & 1.14 & 1.02 & 1.50 \\
        circles & 22.7 & 4.79 & 18.02 & 1.43 \\
        squares & 22.9 & 1.96 & 3.84 & 1.34 \\
        triangles & 23.1 & 2.35 & 2.40 & 1.31 \\
        circles & 23.3 & 2.37 & 4.41 & 1.41 \\
        triangles & 25.9 & 1.08 & 0.50 & 1.33 \\
        circles & 29.3 & 1.15 & 1.04 & 1.57 \\
        circles & 35.6 & 1.13 & 1.01 & 1.45 \\
        circles & 40.7 & 1.11 & 0.96 & 1.51 \\
        triangles & 44.1 & 1.60 & 1.11 & 1.33 \\
        squares & 44.7 & 1.94 & 3.77 & 1.42 \\
        squares & 52.7 & 1.94 & 3.78 & 1.32 \\
        squares & 59.4 & 1.94 & 3.75 & 1.39 \\
  \end{tabular}
  \caption{All experimental geometries used in this research. Ordered by void fraction $\phi$ and showing hole shape; hole characteristic width $W$; hole area $A$; and mean bubble size $\bar{R_0}$.}
  \label{tab:all geometries}
  \end{center}
\end{table}

\subsection{Numerical model}
\label{sec:numerical model}
Bubbles collapsing in complex geometries experience varied degrees of asymmetry. This asymmetry can be quantified with the `anisotropy parameter' $\zeta$, a dimensionless equivalent of the Kelvin impulse, which can predict several bubble collapse properties \citep{Supponen2016, Supponen2017, Supponen2018}.

We have previously presented a numerical model, based on the boundary element method, capable of predicting the anisotropy parameter for arbitrary rigid geometries \citep{Andrews2022}. This model assumes that the bubble is spherical and stationary. Thus, the bubble can be treated as a fixed three-dimensional point sink in potential flow with a strength depending on the bubble radius and radial velocity of the bubble surface which are calculated by numerically solving the Rayleigh-Plesset equation. Nearby boundaries are modelled by a distribution of point sink elements and the no-through-flow boundary condition is imposed at their centroids. Each element $i$ has a constant sink strength density $\sigma_i$ across its surface and area $A_i$ such that the sink strength of the centroid point sink is $\sigma_i A_i$ and the velocity induced at a point $j$ by the element $i$ is
\begin{equation}
    \nabla \Phi |_j = \frac{\sigma_i A_i (\mathbf{x}_j - \mathbf{x}_i)}{4 \pi |\mathbf{x}_j - \mathbf{x}_i|^3},
\end{equation}
where $\Phi$ is the velocity potential, $\mathbf{x}_j$ is a point in the fluid, and $\mathbf{x}_i$ is the centroid position of the element $i$.

The bubble is assumed to be stationary and the integral of pressures on the boundary surface is calculated and integrated over time. This allows the Kelvin impulse to be estimated and then non-dimensionalised to the anisotropy parameter $\zeta$.

Porous plates could be directly modelled this way, however a high number of elements would be required to adequately resolve the holes. Here we propose an adaptation of the previous model that does not require each hole to be resolved separately.

By assuming that the plate has zero thickness, the plate can be modelled with a single layer of boundary elements. Elements could then be constructed to surround the holes, however this would simply act to redistribute the element centroids and reduce the overall area. Instead, we assume that the plate is homogeneous, and simply scale the element areas using the void fraction so that the porous element area $A_i$ is
\begin{equation}
    A_i = (1 - \phi) A_{is}
\end{equation}
where $A_{is}$ is the area of the equivalent non-porous element. This method assumes that the shape of the holes doesn't matter, and that the holes are small enough that any difference in horizontal position is negligible.

After scaling the areas, the boundary element method solution can proceed as usual, using the reduced $A_i$ for both the boundary conditions and pressure integration. We have previously presented the core boundary element method procedure \citep{Andrews2020} and anisotropy computations \citep{Andrews2022} in detail.

\section{Experimental results}
\subsection{Bubbles close to porous boundaries}
Bubbles collapsing in close proximity to porous boundaries show a broad range of interesting dynamics. In this section, frames from high-speed recordings of five bubble collapses are presented, demonstrating some of these dynamics.

\begin{figure}
    \centering
    \includegraphics{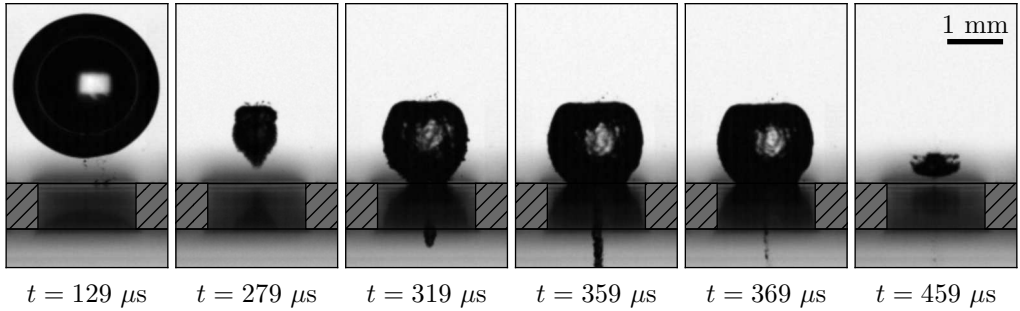}
    \caption{Frames from a high-speed recording of a bubble collapsing with a strong jet passing through a hole in a porous plate with void fraction $\phi = 44.7$ \% and square holes of size $W = 1.94$ mm. A cross-section of the plate at the bubble position is superimposed with hatched areas indicating the solid part of the plate.}
    \label{fig:strong_jet}
\end{figure}

When bubbles collapse close to non-porous boundaries, they produce strong jets that often impinge on the boundaries. When the bubble is nucleated above a hole (positioned above the orange circles in figure \ref{fig:schematic}), these jets can propagate through the hole, producing very long regions of entrained vapour. One such example is shown in figure \ref{fig:strong_jet}. In this figure, by measuring the position of the vapour jet tip, an average jet velocity of 39 m s$^{-1}$ is found between $t = 279$ $\mu$s and $t = 319$ $\mu$s. The frame at $t = 359$ $\mu$s shows that the vapour entrained by the jet extends beyond the bottom of the frame. This region of vapour then itself very rapidly collapses, as seen at $t = 369$ $\mu$s, followed by the rest of the bubble collapsing down towards the hole.

\begin{figure}
    \centering
    \includegraphics{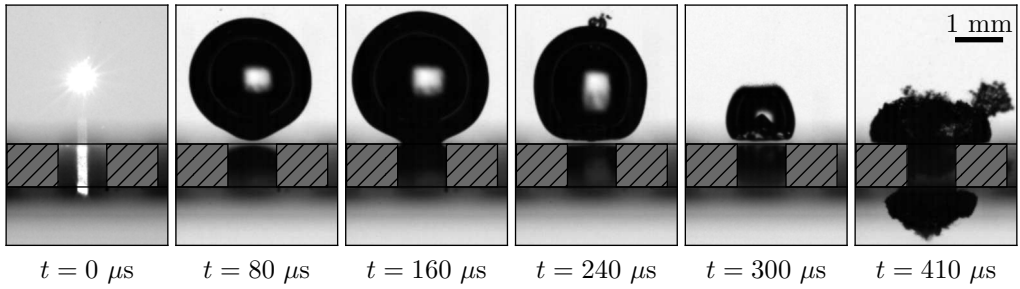}
    \caption{Frames from a high-speed recording of a bubble collapsing through a hole in a porous plate with void fraction $\phi = 21.6$ \% and circular holes of size $W = 1.14$ mm. A cross-section of the plate at the bubble position is superimposed with hatched areas indicating the solid part of the plate.}
    \label{fig:on_hole_inversion}
\end{figure}

As well as allowing jets to pass through the holes, the holes reduce the impedance of the boundary. Bubbles collapse towards rigid boundaries because the rigid boundary impedes the flow out from the bubble during expansion and into the bubble during collapse. This results in the opposite side of the bubble expanding and collapsing much faster, leading to the overall motion towards the boundary \citep{Blake1983}. When the bubble is directly over a hole (positioned above the orange circles in figure \ref{fig:schematic}), the hole does not impede the bubble expansion and collapse. Figure \ref{fig:on_hole_inversion} shows one such configuration. Initially, the bubble expands mostly spherically. However, the bottom of the bubble, positioned directly over the hole, expands more than the rest of the lower side, resulting in a protrusion visible in the frame at $t = 80$ $\mu$s. The protrusion expands into the hole when the bubble is at its maximum size at $t = 160$ $\mu$s. Subsequently, during collapse, the protrusion entirely withdraws by $t = 240$ $\mu$s, with a clear gap visible between the now-flattened lower surface and the boundary. Due to the low impedance of the hole compared to the surrounding boundary, the center of the bottom side of the bubble continues collapsing much more rapidly than the surrounding areas. This leads to a full inversion of the bottom of the bubble, visible by the upward-facing triangular jet entering from the bottom side of the bubble at $t = 300$ $\mu$s. These qualitative observations agree well with those reported by \citet{Khoo2005}. The bubble then fully collapses and re-expands by $t = 410$ $\mu$s. It is interesting to note in the frame at $t = 410$ $\mu$s that the bubble undergoes a rapid ejection event on the upper right surface, the cause of which is unknown. The bubble goes on to travel through the hole, breaking up as it goes. 

\begin{figure}
    \centering
    \includegraphics{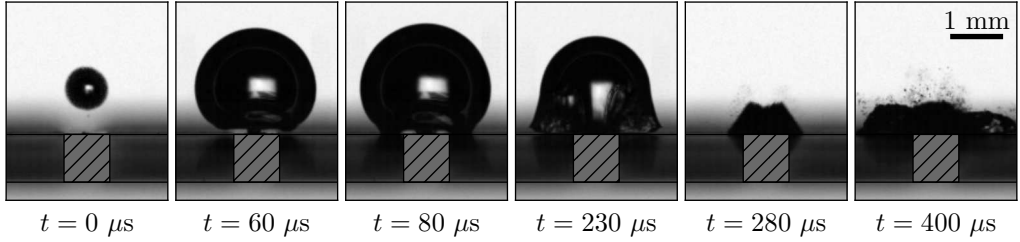}
    \caption{Frames from a high-speed recording of a bubble collapsing above the area between four holes in a porous plate with void fraction $\phi = 44.7$ \% and square holes of size $W = 1.94$ mm. A cross-section of the plate through the row of holes in front of the bubble is superimposed with hatched areas indicating the solid part of the plate.}
    \label{fig:preferential_expansion}
\end{figure}

Bubbles positioned above the solid boundary between holes also expand preferentially towards nearby holes. Figure \ref{fig:preferential_expansion} shows a bubble positioned between four square holes (above the green cross in figure \ref{fig:schematic}\textit{b}). At the lower edge of the bubble, the areas closest to the holes expand more than the center. These appear as two sharper protrusions on either side of the bottom of the bubble in the frames at $t = 60$ $\mu$s and $t = 80$ $\mu$s. On the boundary, an extremely small bubble is nucleated by the laser impinging on the boundary. This bubble is visible in the first three frames as a shadow directly below the main bubble. The main bubble, however, does not contact the boundary at the center until the frame at $t = 230$ $\mu$s, showing how the solid parts of the boundary strongly limit the growth and collapse of the bubble when compared to the holes. As the bubble begins to collapse, surface instabilities become visible in the frame at $t = 230$ $\mu$s, growing initially from the holes and spreading across the bubble surface. The top surface of the bubble rapidly collapses, leaving an indentation on the top of the collapsing bubble at $t = 280$ $\mu$s. The bubble then re-expands along the solid parts of the boundary, splitting into multiple sections visible at $t = 400$ $\mu$s.

\begin{figure}
    \centering
    \includegraphics{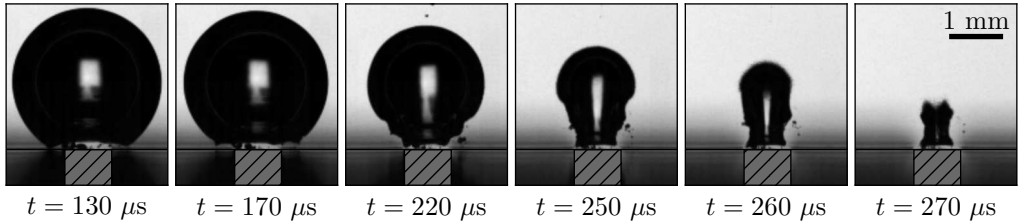}
    \caption{Frames from a high-speed recording of a bubble collapsing above the area between two holes in a porous plate with void fraction $\phi = 44.7$ \% and square holes of size $W = 1.94$ mm. A cross-section of the plate at the bubble position is superimposed with hatched areas indicating the solid part of the plate.}
    \label{fig:collapse_ears}
\end{figure}

These effects can be further explored by observing a bubble positioned between just two holes. This case is shown in figure \ref{fig:collapse_ears}. As in figure \ref{fig:preferential_expansion}, the bubble in figure \ref{fig:collapse_ears} expands preferentially towards the nearby holes. Again, protrusions are visible on either side of the lower surface of the bubble. However, as the bubble collapses, the motion of these protrusions is in the camera plane and so can be more clearly observed. The surface of the bubble directly adjacent to the solid part of the boundary remains significantly impeded throughout the collapse and so does not move significantly. At $t = 170$ $\mu$s the two expanded protrusions have begun to collapse towards bubble centroid. As the collapse advances the protrusions retain their additional curvature, forming `ears' on either side of the bubble which are visible until $t = 270$ $\mu$s. The longevity of these ears is a surprising feature as the areas of a bubble with greatest curvature are expected to collapse most rapidly \citep{Lauterborn1982}.

\begin{figure}
    \centering
    \includegraphics{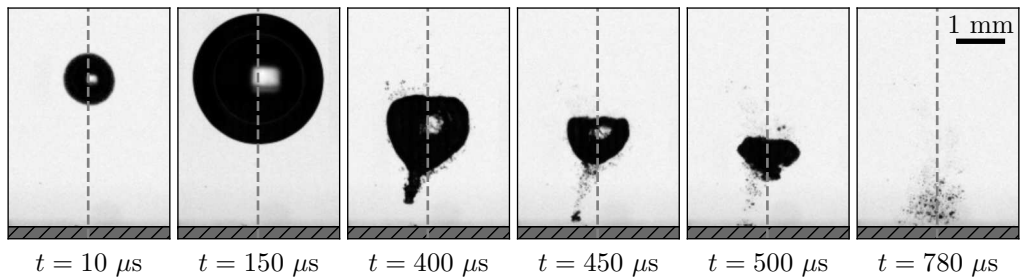}
    \caption{Frames from a high-speed recording of a bubble collapsing above a non-porous boundary. The horizontal position of the bubble at its maximum size is shown by the grey vertical line in each frame. The plate is indicated by the hatched grey area at the bottom of the frame.}
    \label{fig:offset_jet}
\end{figure}

We note here that the rapid jet, often seen around the time of the first re-expansion of the bubble, is not always aligned with the overall motion of the bubble. This suggests that small asymmetries in the initial plasma formation may affect the bubble dynamics beyond the initial formation and expansion, despite the bubble being very spherical at its maximum size. The mechanism for this behaviour could be some history of the initial plasma and expansion being retained due to insufficient mixing and homogenisation of the internal gases of the bubble. An example of an offset jet is shown in figure \ref{fig:offset_jet} where the bubble is collapsing near a non-porous plate. Despite the bubble appearing very spherical at its maximum size, and the plate being highly symmetric, the jet that appears at $t = 400$ $\mu$s is clearly offset from the vertical axis (shown by the grey dashed line). Although this jet is strong, it does not affect the overall motion of the bubble as the bubble proceeds to collapse in the expected vertical direction. This effect is visible in data sets we have previously used \citep{Andrews2022} and can be found in some other publications such as figure 5 in \citet{Pozar2021} and figure 4(\textit{a}) in \citet{Sieber2022}.

\subsection{Variation of displacement and rebound radius with standoff distance}

\begin{figure}
    \centering
    \includegraphics{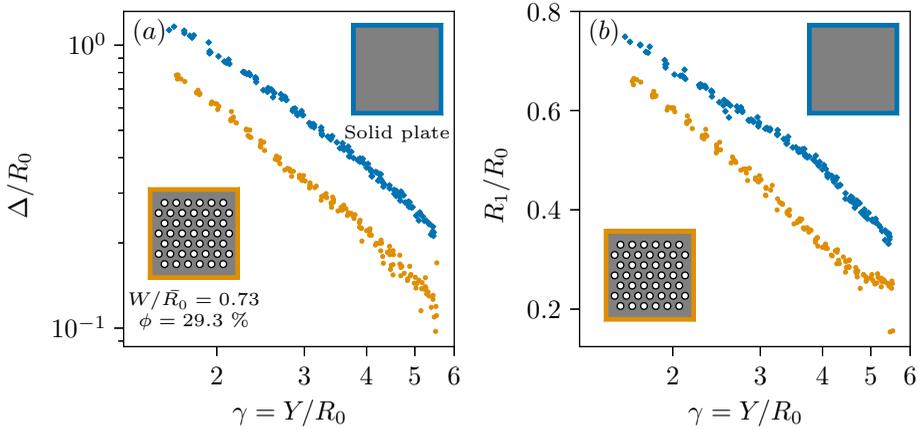}
    \caption{($a$) Normalised displacement plotted against standoff distance. ($b$) Normalised rebound radius plotted against standoff distance. Blue diamonds are data from bubbles collapsing near a non-porous plate. Orange circles are data from bubbles collapsing near a porous plate.}
    \label{fig:simple_svd}
\end{figure}

Bubble displacement $\Delta / R_0$ and rebound radius $R_1 / R_0$ are known to depend strongly on the standoff distance $\gamma = Y / R_0$ \citep{Supponen2016, Supponen2018}. In figure \ref{fig:simple_svd}, the blue data points are experimental data from bubbles collapsing near a non-porous plate. As the standoff distance increases, the displacement decreases, approximately following a power law. Similarly, the rebound radius decreases as the standoff distance increases, although this approximately follows a log law rather than a power law. A porous plate impedes flow, but to a lesser degree than a non-porous plate. Thus, it induces a lesser displacement and rebound radius than a non-porous plate as shown by the orange data in figure \ref{fig:simple_svd}. For the porous plate, the displacement approximately follows a power law and the rebound radius approximately follows a log law, which are the same trends as for the non-porous plate.

\subsection{Comparing horizontal position, hole size, and hole shape}

Many parameters govern the porous plate and a vast number of plates could be considered with a range of hole shapes and sizes. In addition, the bubble can vary in position relative to the plate both horizontally and vertically. In order to reduce the parameter space, it is thus desirable to determine which parameters can be considered negligible.

As discussed in section \ref{sec:problem definition}, the horizontal position of the bubble has two extremes: above a hole (above the orange circles in figure \ref{fig:schematic}), or directly above an area of plate between holes (above the green crosses in figure \ref{fig:schematic}). Intuitively, it can be understood that a bubble above a hole will displace less than a bubble above solid boundary because the fluid does not impede the collapse of the bottom of the bubble. In the limiting case of an infinitely large hole, the bubble experiences no asymmetry, and thus no displacement. Conversely, with a fixed void fraction, infinitely large holes produce infinitely large spaces between holes. Thus, for bubbles nucleated between-holes, the bubble is no longer affected by the holes and so tends towards the solution for a simple non-porous plate. However, for sufficiently small holes, the difference between bubbles collapsing above a hole and bubbles collapsing between-holes is expected to become negligible.

\begin{figure}
    \centering
    \includegraphics{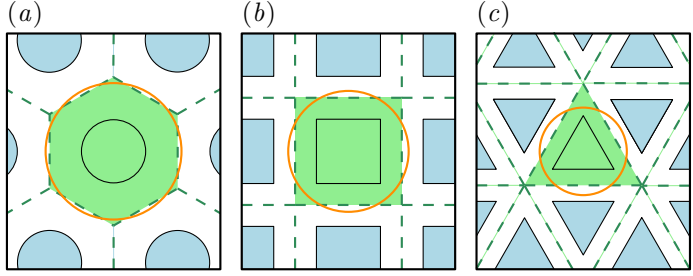}
    \caption{Top-down view diagrams of porous plates with circular holes (\textit{a}), square holes (\textit{b}), and triangular holes (\textit{c}). The tessellation pattern is shown by the green dashed lines with a single tessellation area shaded in green. The orange circles indicate the size of a bubble with equal area to each tessellation area such that $A' = 1$.}
    \label{fig:tessellation}
\end{figure}

Here we define a dimensionless area parameter $A'$ to be the ratio between the area of one tessellation unit and the projected area of the bubble. The tessellation unit area is the total area that would be used to calculate the void fraction of one hole. These units are shown graphically in figure \ref{fig:tessellation}. Thus, the parameter $A'$ is defined as
\begin{equation}
    A' = \frac{A}{\phi \pi \bar{R_0}^2}
\end{equation}
where $A$ is the area of one hole and $\bar{R_0}$ is the mean maximum bubble radius of bubble collapse experiments near the plate.

\begin{figure}
    \centering
    \includegraphics{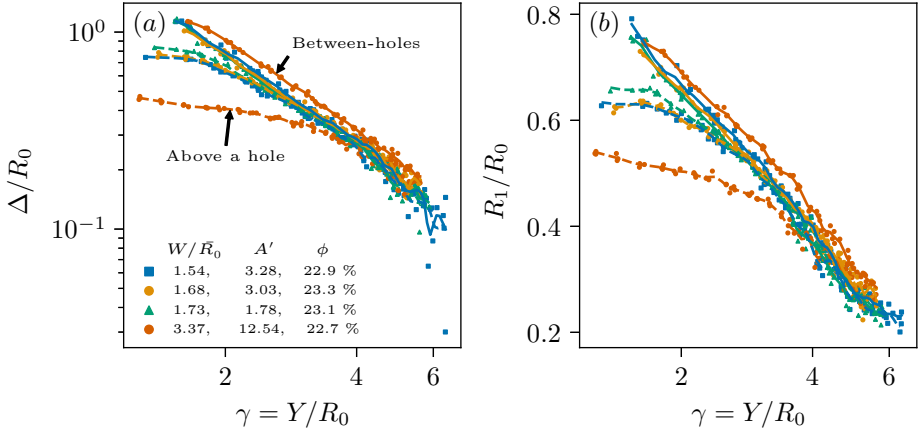}
    \caption{($a$) Normalised displacement plotted against standoff distance. ($b$) Normalised rebound radius plotted against standoff distance. Data are plotted for four porous plates with tesselation unit areas $A' > 1.5$. To more easily distinguish data sets, data for bubbles positioned between-holes are traced by solid lines and data for bubbles positioned above holes are traced by dashed lines. Data markers have shapes corresponding to the shape of holes in the plates (circles, squares, and triangles).}
    \label{fig:svd_position}
\end{figure}

Figure \ref{fig:svd_position} shows data for four different plates. One plate has very large tessellation unit area compared to the average bubble size ($A' = 12.54$), while the other three have area ratios in the range $1.78 \leq A' \leq 3.28$. For each plate, bubbles are positioned above a hole and between-holes. The dashed lines follow bubbles above holes and the solid lines follow bubbles between-holes. It is immediately clear that there is a distinct separation of data, with bubbles above holes displacing significantly less, and rebounding to a smaller size, than bubbles between-holes. However, as the standoff distance increases, the dashed and solid lines converge. For the three plates with smaller holes, at sufficiently large standoff distances, the lines merge completely and the data become independent of horizontal position (within experimental variation). For these three data sets, the data become independent of horizontal position at approximately $\gamma = 3$. In general, as $A'$ decreases, convergence occurs at lower standoff distances.  Thus, the horizontal position of the bubble is unimportant for displacement and rebound size when the pattern of holes is on a scale smaller than the bubble size. This is confirmed by figure \ref{fig:simple_svd} where the porous plate data (with $A' = 0.13$) for above a hole and between-holes are plotted together and show no greater spread than the non-porous plate data in the same figure. In addition, for small holes (low $W / R_0$) at very low void fractions, the dimensionless area $A'$ can be large but does not result in significant splitting of the data.

The data in figure \ref{fig:svd_position} represent three different hole shapes: a square, a circle, and a triangle. The square and smaller circular holes have very similar area ratios and produce almost identical curves for both displacement and rebound size. The triangular holes are slightly smaller and show slightly less splitting of the data. Thus, using the dimensionless area $A'$, the shape of holes can be considered unimportant. This can be explained by the short timescale on which these flows occur. In less rapid flows, the shape of a hole is important due to the viscous boundary layers that form around the edge of the hole. Shapes such as triangles have a greater perimeter per unit area when compared to circles. This increased surface leads to more boundary layers forming which further restrict the flow. However, for flow induced by a bubble collapsing near a hole, there is insufficient time for a significant viscous boundary layer to form and thus the shape of the hole becomes insignificant. This can be shown with the approximate relation $\delta \sim \sqrt{\nu t}$ where $\delta$ is the approximate scale of the boundary layer thickness, $\nu$ is the kinematic viscosity, and $t$ is the time over which the boundary layer would develop. The kinematic viscosity of water at room temperature is approximately $1 \times 10^{-6}$ m$^2$ s$^{-1}$. From our experiments, a typical initial growth and collapse cycle occurs in approximately 0.5 ms. Thus, the boundary layer formed in this time would be on the scale of 0.02 mm which is much smaller than the size of the holes and so can be considered insignificant.

\subsection{Variation of displacement and rebound ratio with void fraction}
From the observations above, we can neglect the influence of hole size, hole shape, and horizontal position for porous plates with dimensionless area $A' < 1.5$ or with small holes ($W / R_0 < 1$). All analysis hereafter relies only on data within this regime and we now further investigate the influence of the void fraction.

\begin{figure}
    \centering
    \includegraphics{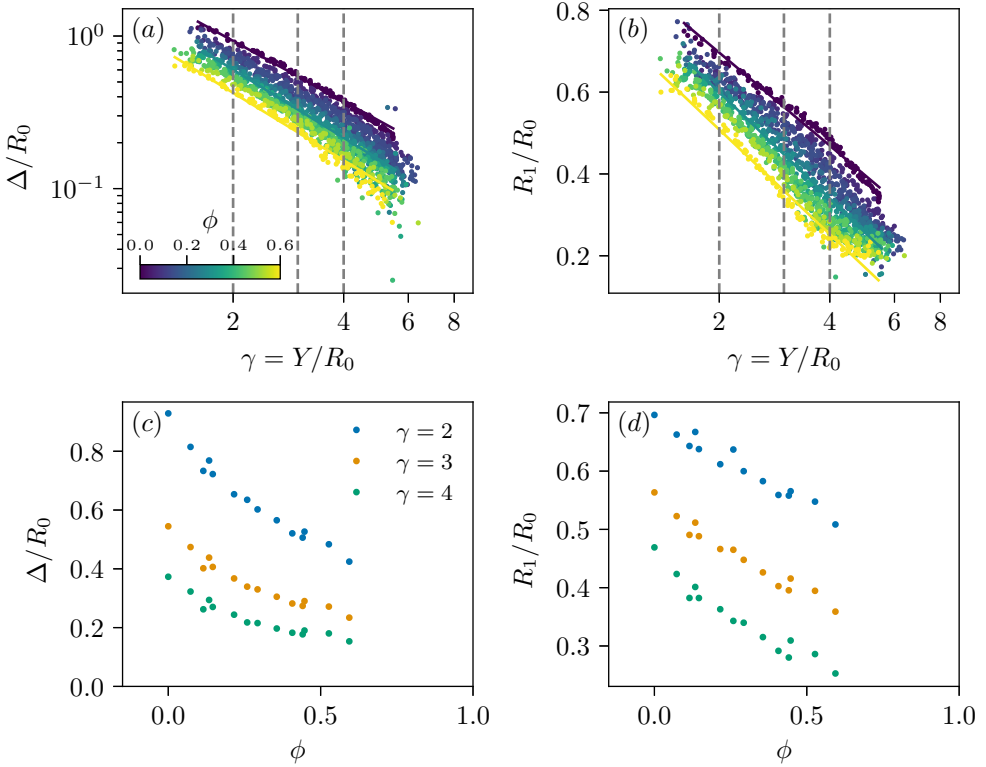}
    \caption{($a$) Normalised displacement plotted against standoff distance for a range of void fractions $\phi$. ($b$) Bubble rebound radius ratio plotted against standoff distance for a range of void fractions $\phi$. Straight-line curve fits are shown for three representative cases in each of ($a$) and ($b$). ($c$) Normalised displacement plotted against void fraction $\phi$ for standoff distances corresponding to the grey dashed lines in ($a$). ($d$) Bubble rebound radius ratio plotted against void fraction for standoff distances corresponding to the grey dashed lines in ($b$). Only data for plates with $A' < 1.5$ or $W / \bar{R_0} < 1$ are included in these plots.}
    \label{fig:svd_vf}
\end{figure}

Starting from a non-porous plate, with zero void fraction, and then increasing the void fraction, figure \ref{fig:svd_vf}(\textit{a}) shows that displacement decreases as void fraction increases. Similarly, figure \ref{fig:svd_vf}(\textit{b}) shows that the rebound size ratio decreases as the void fraction increases. Both of these confirm that higher plate porosity results in less asymmetry in the bubble collapse. Despite the decrease in displacement and rebound ratio with increasing void fraction, the gradient remains remarkably similar for all void fractions.

To find out how the displacement and rebound size depend on void fraction, we take vertical slices at three stand-off distances. For each slice, the displacement and rebound size values at a given void fraction are calculated from straight-line fits to each data set shown in figures \ref{fig:svd_vf}(\textit{a}) and \ref{fig:svd_vf}(\textit{b}). Three such curve fits are shown on each of figures \ref{fig:svd_vf}(\textit{a}) and \ref{fig:svd_vf}(\textit{b}) as examples. Figures \ref{fig:svd_vf}(\textit{c}) and \ref{fig:svd_vf}(\textit{d}) show displacement and rebound size ratio as functions of void fraction for three values of standoff distance. The displacement and rebound radius both decrease as the void fraction increases. Both displacement and rebound radius show fairly uniform sensitivity to void fraction across different standoff distances, although the displacement is marginally more sensitive at low standoff distances.

\section{Anisotropy parameter for porous plates}
In the previous section, we have shown how displacement and radius ratio vary with both standoff distance and void fraction. In order to unify these parameters, and compare these results with other geometries, it is desirable to formulate the anisotropy parameter as a function of standoff distance and void fraction.
We assume that the anisotropy of a porous plate is only a function of void fraction $\phi$ and standoff distance $\gamma$.
\begin{equation}
    \zeta = f(\phi, \gamma)
    \label{eq:f_function}
\end{equation}
In this section we present two methods of determining the function $f(\phi, \gamma)$.

\subsection{Displacement and rebound ratio as a function of anisotropy}
\begin{figure}
    \centering
    \includegraphics{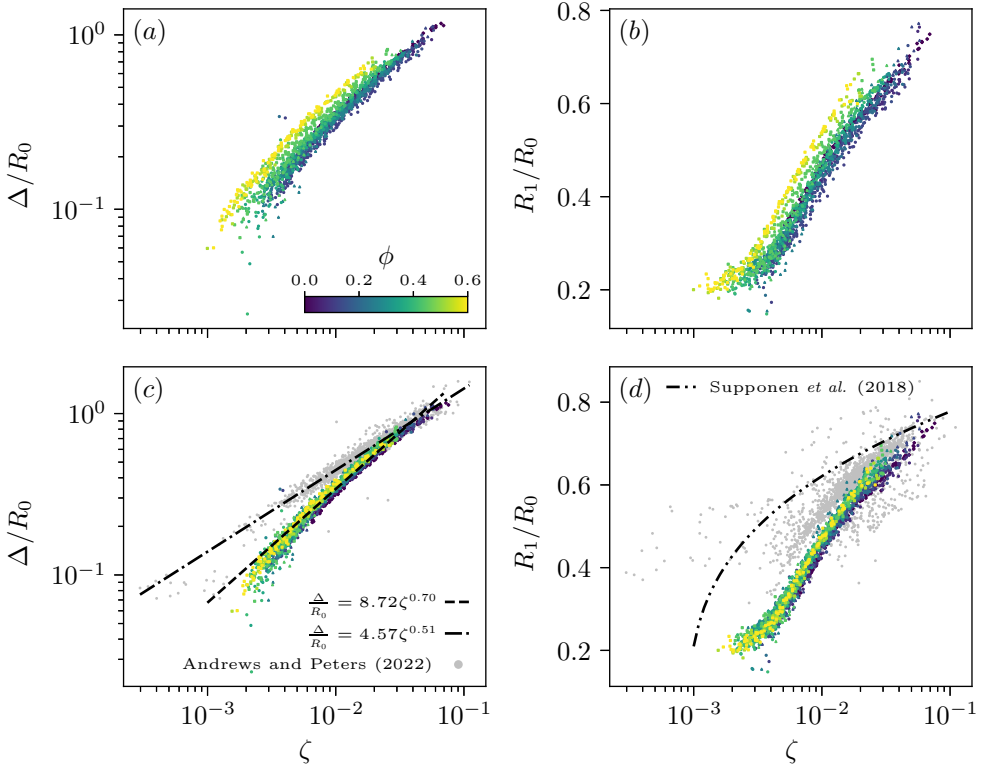}
    \caption{Normalised displacement ($a$) and rebound radius ratio ($b$) plotted against the anisotropy parameter magnitude $\zeta$ predicted by the numerical model. Normalised displacement ($c$) and rebound radius ratio ($d$) plotted against the anisotropy parameter magnitude $\zeta$ estimated by fitting displacement data to the non-porous plate data. Data are coloured by void fraction $\phi$. The grey data points in ($c$) and ($d$) are from \citet{Andrews2022}. The black dashed line is a curve fit to the porous plates data. The black dash-dotted line is the curve fit from \citet{Andrews2022}. The black dash-dot-dotted line is derived from the curve fit of \citet{Supponen2018}.}
    \label{fig:all_anisotropy_collapse}
\end{figure}

The first formulation of the anisotropy parameter is the implementation of the numerical method described in section \ref{sec:numerical model}. Using this method, experimental measurements of displacement and rebound ratio can be plotted as a function of the anisotropy predictions as shown in figure \ref{fig:all_anisotropy_collapse} (parts \textit{a} and \textit{b}). Although not perfect, it shows reasonable collapse of most data onto a single curve for each of the two measurements. 

\begin{figure}
    \centering
    \includegraphics{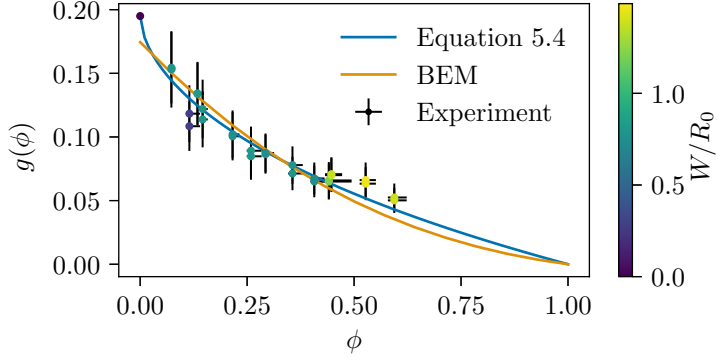}
    \caption{Prefactors $g(\phi)$ where $\zeta = g(\phi) \gamma^{-2}$ plotted against void fraction $\phi$. Experiment data points are computed from fitting displacement data for porous plates with the non-porous plate data. Horizontal error bars represent the range of possible void fractions for each data point. Vertical error bars are the standard deviation from the least squares fit of $g(\phi)$. Points are coloured by the dimensionless width of the holes.}
    \label{fig:c_vs_phi}
\end{figure}

Using the numerical model, we find that the anisotropy parameter varies almost exactly with $\gamma^{-2}$ across all porous plates. This is consistent with all the other anisotropy functions for flat geometries presented by \citet{Supponen2016}. Thus, we simplify equation \ref{eq:f_function} to
\begin{equation}
    \zeta = g(\phi)\gamma^{-2},
\label{eq:g_function}
\end{equation}
leaving only the function $g(\phi)$ to be determined. Using the numerical model, the prefactor $g(\phi)$ is plotted as the orange line in figure \ref{fig:c_vs_phi}. This was computed for a $50$ mm $\times$ $50$ mm plate using 19 756 elements, with element lengths ranging between $0.35$ mm and $0.42$ mm, where larger elements were used around the edge of the plate. As the void fraction increases, the prefactor $g(\phi)$ decreases. At a void fraction $\phi = 0$, identically a non-porous plate, the prefactor approaches the solution for a non-porous plate $g(0) = 0.195$. Notably, due to differences between the numerical model and analytic solution for a flat plate, the boundary element method solution does not reach $0.195$. At the opposite limit, with void fraction $\phi = 1$, the plate does not exist, resulting in zero anisotropy, thus $g(1) = 0$. Between these limits the gradient of the prefactor is highest at low void fractions, indicating higher sensitivity to void fraction when the void fraction is low. This conclusion is also reflected in the shape of the curves in figure \ref{fig:svd_vf}(\textit{c}, \textit{d}).

The second formulation for the anisotropy parameter assumes that displacement is solely a function of the anisotropy parameter $\zeta$ and maintains the assumption that the anisotropy can be written as equation \ref{eq:g_function}. The analytic solution for a non-porous plate ($g(0) = 0.195$) can be applied to the non-porous plate data to give the measured displacement as a function of anisotropy. Then, for each porous plate data set, the prefactor $g(\phi)$ can be fitted such that the porous plate data follows the same curve for displacement against anisotropy. This curve fit is performed using the logarithmic least-squares difference between each data set and the curve fit on the non-porous plate data.

Using the fitted prefactors $g(\phi)$, the data collapses very well onto single curves for displacement and rebound radius as shown in figure \ref{fig:all_anisotropy_collapse} parts (\textit{c}) and (\textit{d}). It should be noted that only the displacement curve is fitted and the resulting values cause the rebound radius data to collapse as well, suggesting that these values are representative of the underlying physics and not simply overfitting to the data.

The fitted prefactors are shown alongside the numerically predicted curve in figure \ref{fig:c_vs_phi} with vertical error bars of one standard deviation of the least-squares fit. In this plot, different horizontal positions are treated as distinct data sets, resulting in two data points for each porous plate which are typically very closely aligned.

Figure \ref{fig:c_vs_phi} shows that the boundary element method agrees well with the experimentally determined prefactors. However, at low void fractions the experimental values tend to be below the numerical predictions, whereas at higher void fractions they tend to be above. This may suggest that there are some nuances to the behaviour that the numerical model cannot capture. For example, the model does not account for how bubble deformation and displacement may interact with the porous plates.

An empirical curve fit for the anisotropy parameter is desirable in order to further reduce the cost of modelling porous boundaries. It is noted that the two limits of $g(\phi)$ are $g(0) = 0.195$ and $g(1) = 0$. We assume that $g(\phi)$ is a non-linear, smooth function between these limits of the form 
\begin{equation}
    g(\phi) = 0.195 (1 - \phi^k)
\end{equation}
where $k$ is the single parameter to be fitted. Using the experimental data shown in figure \ref{fig:c_vs_phi}, the fitted parameter $k$ is found to be 0.50 with a standard deviation of 0.03. This curve fit is plotted alongside the data in figure \ref{fig:c_vs_phi} which shows good agreement with experimental data. The anisotropy can therefore be written as
\begin{equation}
    \zeta = 0.195 (1 - \phi^{0.5})\gamma^{-2}
    \label{eq:anisotropy equation}
\end{equation}
for all $\phi \in [0, 1]$, effectively providing a single equation for displacement and rebound radius for any void fraction and standoff distance.

\subsection{Disparity with other experimental methodologies}
Computing the anisotropy for this data allows it to be compared to prior research, as shown in figure \ref{fig:all_anisotropy_collapse} (parts \textit{c} and \textit{d}). Figure \ref{fig:all_anisotropy_collapse}(\textit{d}) shows the curve presented by \citet{Supponen2017} as well as data points from \citet{Andrews2022} which used a different experimental method than the present research. Although there is significant spread in the data from \citet{Andrews2022}, it does not follow the same curve as the present research. The curve presented by \citet{Supponen2017} is even further different. This reinforces our previous suggestion that there is likely another factor that varies between experimental methodologies that can significantly affect the bubble rebound size \citep{Andrews2022}.

Figure \ref{fig:all_anisotropy_collapse}(\textit{c}) shows data points and the curve fit for displacement against anisotropy for a range of complex geometries presented by \citet{Andrews2022}. This collapsed curve is markedly different to the curve presented in the present research. The principal difference between the two works is that the previous research used a microscope objective to create bubbles whereas the current research uses an off-axis parabolic mirror. This difference in displacement is likely partly due to smaller rebounds, but may also be affected by bubble morphology due to the difference in plasma shapes created by different focusing optical elements.

\section{Conclusion}
In this research, we have investigated how a pattern of through-holes in a rigid boundary affect the dynamics of a collapsing bubble. We have demonstrated how bubbles expand preferentially towards the holes and less towards the solid parts of the boundaries. We have shown that the displacement and rebound radius do not depend significantly on the shape of the holes, and the size of the holes only becomes important when comparable to the bubble size (dimensionless tessellation unit area $A' > 1.5$). Bubbles used in ultrasonic cleaning theoretically vary between 0.1 $\mu$m and 100 $\mu$m in radius, depending on the driving frequency and power, with experiments showing typical radii in the order of a few $\mu$m up to approximately $20~\mu$m \citep{Brotchie2009, FernandezRivas2012, FernandezRivas2013a}. The pore size of filters varies depending on the intended application. \cite{Reuter2017} reference a pore size of 30 nm, for example. Thus, the applicability of this regime varies with application.

The bubble displacement and rebound radius depend strongly on both the standoff distance and void fraction of the porous plate. These parameters can be unified in terms of the anisotropy parameter with equation \ref{eq:anisotropy equation}. Using this unified parameter, all data for porous plates collapse onto single curves for displacement and rebound radius. However, the collapsed curves vary from those found for our previous work \citep{Andrews2022} which used a different experimental method. Nevertheless, equation \ref{eq:anisotropy equation} can be combined with scaling laws from prior research \citep{Supponen2016} to predict other bubble collapse properties such as jet speed.

This work provides a solid first step towards characterising bubble behaviour near porous plates and connects this geometry to the wider framework of investigations using the anisotropy parameter. Further work is required to connect this framework of single-bubble collapse to applications such as ultrasonic cleaning. For example, understanding the geometric distribution of bubble collapse events induced by an ultrasound field; the combined effect of multiple bubbles; and the relation between surface shear stress and the anisotropy parameter. Investigation of the surface pressure distribution of porous plates, and complex geometries in general, would also provide valuable insight into the cleaning effects of collapsing bubbles.

\section*{Acknowledgements}
We acknowledge financial support from the EPSRC under Grant No. EP/P012981/1. DFR acknowledges the funding from the European Research Council (ERC) under the European Union's Horizon 2020 research and innovation programme (Grant agreement No. 851630).

\section*{Declaration of interests}
The authors report no conflict of interest.

\bibliographystyle{jfm}

\end{document}